\documentclass[aps,prd,preprint,showpacs,superscriptaddress]{revtex4-1}
\usepackage{amsmath}
\usepackage{latexsym}
\usepackage{amsfonts}
\usepackage{graphicx}
\usepackage{mathrsfs}
\usepackage{epstopdf}

\begin{document}

\title{Inflationary models with non-minimally derivative coupling}

\author{Nan Yang}
\email{cqunanyang@hotmail.com}
\affiliation{School of Physics, Huazhong University of Science and Technology,
Wuhan, Hubei 430074, China}

\author{Qin Fei}
\email{feiqin@hust.edu.cn}
\affiliation{School of Physics, Huazhong University of Science and Technology,
Wuhan, Hubei 430074, China}

\author{Qing Gao}
\email{gaoqing1024@swu.edu.cn}
\affiliation{School of Physical Science and Technology, Southwest University, Chongqing 400715, China}

\author{Yungui Gong}
\email{yggong@hust.edu.cn}
\affiliation{School of Physics, Huazhong University of Science and Technology,
Wuhan, Hubei 430074, China}
\affiliation{CASPER, Department of Physics, Baylor University, Waco, Texas 76798, USA}

\date{\today}

\begin{abstract}
We derive the general formulae for the the scalar and tensor spectral tilts to the second order
for the inflationary models with non-minimally derivative coupling without taking the high friction limit. The non-minimally
kinetic coupling to Einstein tensor brings the energy scale in the inflationary models down to be sub-Planckian.
In the high friction limit, the Lyth bound is modified with an extra suppression factor, so that
the field excursion of the inflaton is sub-Planckian. The inflationary models with
non-minimally derivative coupling are more
consistent with observations in the high friction limit. In particular, with
the help of the non-minimally derivative coupling, the quartic power law potential is consistent with the observational
constraint at 95\% CL.
\end{abstract}

\preprint{1504.05839}

\maketitle
\section{Introduction}

Inflation successfully solves various problems in the standard big bang cosmology
such as the flatness, horizon and monopole problems, etc, and the quantum
fluctuation of the inflaton seeds the formation of large-scale structure \cite{starobinskyfr, guth81, linde83, Albrecht:1982wi}.
A scalar field with a flat potential is usually used to model inflation.
The potential of the scalar field is characterized by the slow-roll parameters,
the spectral tilts and the tensor to scalar ratio are approximated by the slow-roll parameters
at the horizon exit. The Planck temperature and polarization data on the measurements of the cosmic microwave background
anisotropies gives $n_s=0.968\pm 0.006$ and $r_{0.002}<0.11$ (95\% CL) \cite{Adam:2015rua,Ade:2015lrj}.
The results are consistent with the $R^2$ inflation \cite{starobinskyfr} and the non-minimally coupled
models at strong coupling limit \cite{Kaiser:1994vs,Bezrukov:2007ep,Kallosh:2013tua}, because both models predict that $n_s=1-2/N$ and $r=12/N^2$,
where $N$ is the number of e-folds before the end of inflation,
but the minimally coupled power law potentials with $n>2$ are disfavored. With the presence of the
non-minimal coupling $\xi R\phi^2$,
the $\lambda\phi^4$ potential with $\lambda\sim O(1)$ can be consistent with the observations \cite{Bezrukov:2010jz}.
Furthermore, the coupling constant $\xi$ can even be as small as $0.003$ \cite{Bezrukov:2013fca,Boubekeur:2015xza}.
If the kinetic term
of the scalar field is non-minimally coupled to Einstein tensor, then
the effective self-coupling $\lambda$ of the Higgs boson can be lowered to be the order of 1, and the new Higgs
inflation introduces no new degree of freedom \cite{Germani:2010gm,Germani:2014hqa}.

The non-minimal coupling can be generalized to be $f(\phi)R$
which is a special case of the general scalar-tensor theory $F(\phi,R)$ \cite{Guo:2010jr,Jiang:2013gza,Boubekeur:2015xza},
because the non-minimal coupling term can be transformed away by a conformal transformation.
If the kinetic term of the scalar field
is non-minimally coupled to curvature tensors, then conformal transformation
cannot transform the model to scalar-tensor theory \cite{Amendola:1993uh}.
More general non-minimally derivative couplings for the scalar field
are discussed in \cite{Horndeski:1974wa,Amendola:1993uh,Capozziello:1999uwa,Capozziello:1999xt}.
The non-minimally derivative coupling usually introduces higher than second order derivatives
in the field equation and more
degrees of freedom, which
lead to the Boulware-Deser ghost \cite{Boulware:1972zf}.
However, Horndeski derived a general scalar-tensor theory with
field equations which are at most of second order in the derivatives of both the metric
$g_{\mu\nu}$ and the scalar field $\phi$ in four dimensions \cite{Horndeski:1974wa}.
In Horndeski theory, the second derivative $\phi_{;\mu\nu}$ couples to Einstein tensor by the
general form $f(\phi,X)G^{\mu\nu}\phi_{;\mu\nu}$, where $X=g^{\mu\nu}\phi_{,\mu}\phi_{,\nu}$ \cite{Horndeski:1974wa}.
If we take $f(\phi,X)=\phi$, then we get the coupling $G^{\mu\nu}\phi_{,\mu}\phi_{,\nu}$ after
integration by parts.
The general derivative coupling which is quadratic in $\phi$ and
linear in $R$, has the forms $\phi_{,\mu}\phi^{,\mu}R$, $\phi_{,\mu}\phi_{,\nu}R^{\mu\nu}$,
$\phi\Box\phi R$, $\phi\phi_{;\mu\nu} R^{\mu\nu}$, $\phi\phi_{,\mu}R^{;\mu}$ and $\phi^2\Box R$.
Due to the divergencies $(R\phi^{,\mu}\phi)_{;\mu}$, $(R^{\mu\nu}\phi\phi_{,\mu})_{;\mu}$ and $(R^{,\mu}\phi^2)_{;\mu}$,
only the couplings $\phi_{,\mu}\phi^{,\mu}R$, $\phi_{,\mu}\phi_{,\nu}R^{\mu\nu}$, and
$\phi\Box\phi R$ are independent.
If we choose the non-minimally derivative coupling as $G^{\mu\nu}\phi_{,\mu}\phi_{,\nu}$,
then the field equations contain no more than second derivatives \cite{Sushkov:2009hk}£¬
and the gravitationally enhanced friction causes the scalar field to evolve more slowly.
For a massless scalar field without the canonical kinetic term $g^{\mu\nu}\phi_{,\mu}\phi_{,\nu}$, the non-minimally
derivative coupled scalar field behaves as a dark matter \cite{Gao:2010vr,Ghalee:2013smy}.
The cosmological perturbations and the first order approximation of the power spectrum for inflationary
models with this non-minimally derivative coupling in the high friction limit
were discussed in \cite{Germani:2010ux,Germani:2011ua,Tsujikawa:2012mk,Sadjadi:2013psa}. As we show above,
the Planck data gives $n_s=0.968\pm 0.006$, so the first order slow-roll parameters are in the order of $0.01$
and the first order approximation is enough. Future
experiments will measure $n_s$ more accurately, so it is necessary to consider the second order corrections. In
this paper, we will derive the first order correction to the amplitude of the power spectrum, the second
order corrections to both the scalar and tensor tilts and the scalar to tensor ratio $r$ without taking the high friction limit.
The cosmological consequences of the theory with non-minimally derivative coupling were also discussed extensively \cite{Daniel:2007kk,Saridakis:2010mf,Sushkov:2012za,Skugoreva:2013ooa,
DeFelice:2011uc,Sadjadi:2010bz,Sadjadi:2013uza,Minamitsuji:2013ura,Granda:2009fh,
Granda:2010hb,Granda:2011eh,deRham:2011by,Jinno:2013fka,Sami:2012uh,Anabalon:2013oea,Rinaldi:2012vy,Koutsoumbas:2013boa,
Cisterna:2014nua,Huang:2014awa,Bravo-Gaete:2013dca,Bravo-Gaete:2014haa,Bruneton:2012zk,Feng:2013pba,
Feng:2014tka,Heisenberg:2014kea,Goodarzi:2014fna,Huang:2015yva,Cisterna:2015yla,Dalianis:2014sqa,Dalianis:2014nwa,Dalianis:2015aba,Ema:2015oaa,Aoki:2015eba,Yang:2015zgh,Zhu:2015lry,Huang:2016qkd}.

In this paper, we discuss the inflationary models with the non-minimally derivative coupling $G^{\mu\nu}\phi_{,\mu}\phi_{,\nu}$.
The paper is organized as follows. In section II, we review the cosmological equations
and the slow-roll approximation for this theory. The general formulae for the power spectrum and the second order corrections
to the scalar and tensor spectral tilts are obtained in section III without taking the high friction limit. In section IV, we consider the power law potential, the hilltop
potential, a simple symmetry breaking potential and the natural inflation by taking the high friction limit, and conclusions are drawn in section V.

\section{The background evolution}

The action for the scalar field with the kinetic term non-minimally coupled to Einstein tensor is
\begin{equation}
\label{eq1}
S = \frac{1}{2}\int {{d^4}x\sqrt { - g} \left[ {M_{pl}^2R - {g^{\mu \nu }}{\partial _\mu }\phi {\partial _\nu }\phi  + \frac{1}{{{M^2}}}{G^{\mu \nu }}{\partial _\mu }\phi {\partial _\nu }\phi  - 2V\left( \phi  \right)} \right]} ,
\end{equation}
where $M_{pl}^2 = (8\pi G)^{-1} $ and $M$ is the coupling constant with the dimension of mass. For convenience, we use
the Arnowitt-Deser-Misner (ADM) formalism \cite{adm}, and the metric is expressed as
\begin{equation}
\label{eq2}
d{s^2} =  - {N^2}d{t^2} + {h_{ij}}\left( {d{x^i} + {N^i}dt} \right)\left( {d{x^j} + {N^j}dt} \right),
\end{equation}
where $N$, $N^i$, $h_{ij}$ are the lapse function, the shift function and the metric for the three dimensional space, respectively.
By using the ADM splitting of space-time, the action (\ref{eq1}) becomes
\begin{equation}
\label{eq3}
\begin{split}
S =& \frac{1}{2}\int dt d^3 x  \left\{ M_{pl}^2\sqrt{h} \left[\,^{(3)} R\left( N + \frac{\dot \phi^2}{2N M^2 M_{pl}^2} \right)
+ \frac{\dot \phi^2}{NM_{pl}^2} - \frac{2NV}{M_{pl}^2}\right. \right.\\
&\left.\left. +( E_{ij} E^{ij} - E^2 )\left(\frac{1}{N} - \frac{\dot \phi^2}{2N^3M^2M_{pl}^2}\right) \right] \right\},
\end{split}
\end{equation}
where
\begin{equation}
\label{eq4}
{E_{ij}}=\frac{1}{2}\left( {{{\dot h}_{ij}} - {\nabla _i}{N_j} - {\nabla _j}{N_i}} \right),
\end{equation}
$E=h^{ij}E_{ij}$, the extrinsic curvature $K_{ij}=E_{ij}/N$, $\dot\phi=d\phi/dt$,
the covariant derivative is with respect to the three dimensional spatial metric $h_{ij}$,
and all the spatial indices are raised
and lowered by the metric $h_{ij}$. Since the lapse and shift function $N$ and $N_i$
contain no time derivative, the variations with respect to them give the
corresponding Hamiltonian and momentum constraints,
\begin{equation}
\label{eq15}
^{(3)}\!R\,\left(N^2 - \frac{\dot \phi^2}{2 M^2 M_{pl}^2}\right) -
({E_{ij}}{E^{ij}} - {E^2})\left(1 - \frac{3\dot \phi^2}{2 N^2 M^2 M_{pl}^2}\right) - \frac{\dot \phi^2}{M_{pl}^2} - \frac{{2{N^2}V\left( \phi  \right)}}{{M_{pl}^2}} = 0,
\end{equation}
\begin{equation}
\label{eq14}
{\nabla_i}\left[ \left(\frac{1}{N} - \frac{{{{\dot \phi }^2}}}{{2{N^3}{M^2}M_{pl}^2}}\right)(E_j^i - \delta_j^iE) \right] = 0.
\end{equation}

For the background with the homogeneous and isotropic flat Friedmann--Robertson--Walker (FRW) metric,
$N=1$, $N_i=0$ and $h_{ij}=a^2\delta_{ij}$, the Hamiltonian constraint (\ref{eq15}) gives the Friedmann equation
\begin{equation}
\label{eq5}
H^2=\left(\frac{\dot a}{a}\right)^2 = \frac{1}{3M_{pl}^2}\left[\frac{\dot \phi^2}{2}\left(1  + 9\frac{H^2}{M^2} \right) + V(\phi) \right],
\end{equation}
and the momentum constraint is satisfied automatically. The equation of motion of the scalar field $\phi$ is
\begin{equation}
\label{eq6}
\frac{d}{{dt}}\left[ {{a^3}\dot \phi \left( {1  + 3\frac{{{H^2}}}{{{M^2}}}} \right)} \right] =  - {a^3}\frac{{dV}}{{d\phi }}.
\end{equation}
In the cosmological background, the non-minimally derivative coupling $G^{\mu\nu}\phi_{,\mu}\phi_{,\nu}/M^2$ becomes $H^2\dot\phi^2/M^2$ which
enhances the friction of the expansion. In the limit $M\rightarrow \infty$, the effect of the non-minimally derivative coupling is negligible,
Eqs. (\ref{eq5}) and (\ref{eq6}) reduce to the standard cosmological equations.
Combining Eqs. (\ref{eq5}) and (\ref{eq6}), we get the Raychaudhuri equation,
\begin{equation}
\label{eq7}
\dot H - \frac{\dot\phi \ddot\phi}{M^2 M_{pl}^2}H = \frac{1}{2}\frac{\dot \phi^2}{M^2 M_{pl}^2}\dot H -
\frac{3}{2}\frac{\dot \phi^2}{M^2 M_{pl}^2} H^2 - \frac{1}{2} \frac{\dot \phi^2}{M_{pl}^2}.
\end{equation}

If the scalar field slowly rolls down the potential, we have the slow-roll conditions,
\begin{equation}
\label{slrcond1}
\begin{split}
\frac{1}{2}\left(1+\frac{9H^2}{M^2}\right)\dot\phi^2\ll V(\phi),\\
|\ddot \phi|\ll |3H\dot\phi|,\\
\left|\frac{2\dot H}{M^2+3H^2}\right|\ll 1.
\end{split}
\end{equation}
Under these slow-roll conditions, Eqs. (\ref{eq5}) and (\ref{eq6}) can be approximated as
\begin{gather}
\label{eq9}
H^2 \approx \frac{V(\phi)}{3M_{pl}^2},\\
\label{eq10}
3 H\dot\phi \left(1+\frac{3H^2}{M^2}\right)  \approx  - V_\phi,
\end{gather}
where $V_\phi=dV/d\phi$. To quantify those slow-roll conditions \eqref{slrcond1}, we introduce the following slow-roll parameters
\begin{equation}
\label{slrparm1}
\begin{split}
\epsilon_v=\frac{M^2_{pl}}{2}\left(\frac{V_\phi}{V} \right)^2\frac{1+9F}{(1+3F)^2},\\
\eta_v=\frac{M^2_{pl}}{1+3F}\frac{V_{\phi\phi}}{V},\\
\xi^2_v=\frac{M^4_{pl}}{(1+3F)^2}\frac{V_\phi V_{\phi\phi\phi}}{V^2},
\end{split}
\end{equation}
where $F=H^2/M^2$. The normalization of $\epsilon_v$ is chosen to recover the definition in Einstein's
general relativity (GR) in the limit $F\ll 1$. By using Eqs. \eqref{eq9} and \eqref{eq10}, we get
\begin{equation}
\label{selfcond1}
\frac{\dot\phi^2(1+9F)}{2V(\phi)}\approx \frac{1}{3}\epsilon_v,
\end{equation}
and
\begin{equation}
\label{selfcond2}
\eta_H=\frac{\ddot\phi}{H\dot\phi}\approx \epsilon_v-\eta_v,
\end{equation}
so $\epsilon_v\ll 1$ and $|\eta_v|\ll 1$ guarantee the satisfaction of the slow-roll conditions.
To the second order of slow-roll approximation, we get
\begin{equation}
\label{selfcond3}
\epsilon_H=-\frac{\dot H}{H^2}\approx \frac{1+3F}{1+9F}\epsilon_v\left(1-\frac{4+39F+117F^2}{3(1+9F)(1+3F)}\epsilon_v+\frac{2(1+6F)}{3(1+3F)}\eta_v\right).
\end{equation}
In the GR limit, $F\ll 1$, we recover the result $\epsilon_H=\epsilon_v(1-4\epsilon_v/3+2\eta_v/3)$.
Using the equations of motion \eqref{eq5}-\eqref{eq7}, we get
\begin{equation}\label{depvdt}
\begin{split}
\dot\epsilon_v&=2H\epsilon_v\left[\frac{2+21F+81F^2}{(1+9F)^2}\epsilon_v-\eta_v-\frac{4+72F+603F^2+2538F^3+5103F^4}{3(1+3F)(1+9F)^3}\epsilon_v^2\right.\\
&\left. \qquad\qquad +\frac{2(2+48F+441F^2+1944F^3+3645F^4}{3(1+3F)(1+9F)^3}\epsilon_v\eta_v-\frac{1}{3}\eta_v^2\right]
\end{split}
\end{equation}
\begin{equation}\label{detvdt}
\dot\eta_v=H\left(\frac{2(1+6F)}{1+9F}\epsilon_v\eta_v-\xi_v^2\right).
\end{equation}

If we further take the high friction limit $H^2 \gg M^2$, i.e., $F\gg 1$, then the slow-roll parameters become
\begin{gather}
\label{eq11a}
\epsilon_v=\frac{M_{pl}^2}{2}\left(\frac{V_\phi}{V} \right)^2\frac{M^2}{H^2},\\
\label{eq12a}
\eta_v = M_{pl}^2\frac{V_{\phi\phi}}{V}\frac{M^2}{3 H^2},\\
\label{xi2eq1}
\xi^2_v  = \left(\frac{M^2}{3H^2}\right)^2M_{pl}^4\frac{V_\phi V_{\phi\phi\phi}}{V^2}.
\end{gather}
To be consistent with the results obtained in \cite{Germani:2010ux},
we define $\epsilon=\epsilon_v/3$, $\eta=\eta_v$ and $\xi^2=\xi_v$ in the high friction limit $F\gg 1$,
and the different slow-roll parameters satisfy the following relations,
\begin{equation}
\label{eq12b}
\epsilon_H\approx \epsilon\left(1-\frac{13}{3}\epsilon+\frac{4}{3}\eta\right), \quad \eta_H\approx 3\epsilon-\eta.
\end{equation}
In the high friction limit, Eqs. \eqref{depvdt} and \eqref{detvdt} become
\begin{gather}
\label{eq38}
\dot \epsilon\approx 2H\epsilon\left(3\epsilon - \eta-7\epsilon^2+\frac{10}{3}\epsilon\eta-\frac{1}{3}\eta^2\right),\\
\label{eq35}
\dot\eta\approx H(4\epsilon\eta-\xi^2).
\end{gather}

Comparing with the slow-roll parameters in GR,
the slow-roll parameters defined in the non-minimally derivative coupling case have
an extra factor $M^2/(3H^2)$ which is small in the high friction limit, so
more potentials can satisfy the slow-roll conditions and inflation can happen more easily in this theory.
In the high friction limit, the number of e-folds before the end of inflation is
\begin{equation}
\label{lythbld}
N(\phi_*)=\int_{\phi_*}^{\phi_e}\frac{H}{\dot\phi}d\phi=\int_{\phi_*}^{\phi_e}\frac{1}{\sqrt{2\epsilon}}\sqrt{\frac{3H^2}{M^2}}\frac{d\phi}{M_{pl}}
>\sqrt{\frac{3}{2}}\frac{H(\phi_e)}{M}\frac{\phi_e-\phi_*}{M_{pl}}.
\end{equation}
So we get an upper bound for the field excursion,
\begin{equation}
\label{lythbld2}
\frac{\Delta\phi}{M_{pl}}<\sqrt{\frac{2}{3}}\, N \frac{M}{H(\phi_e)}.
\end{equation}
In the high friction limit, $M\ll H$, the field
excursion can be sub-Planckian if $M/H(\phi_e)<N^{-1}$. If $\epsilon(\phi)$ increases
with time, then we can derive a lower bound,
\begin{equation}
\label{lythbld3}
\frac{\Delta\phi}{M_{pl}}>N \frac{M}{H(\phi_*)}\sqrt{\frac{2\epsilon(\phi_*)}{3}}.
\end{equation}
So the Lyth bound in inflationary models with non-minimally derivative coupling becomes
\begin{equation}
\label{lythbld1}
\sqrt{\frac{2}{3}}\, N \frac{M}{H(\phi_e)}>\frac{\Delta\phi}{M_{pl}}>N \frac{M}{H(\phi_*)}\sqrt{\frac{2\epsilon(\phi_*)}{3}}.
\end{equation}
If $\epsilon$ is not a monotonic function, then the Lyth bound in GR can be modified to be smaller \cite{Gao:2014pca}.

The non-minimally derivative coupling introduces the energy scale $M$, one may wonder whether it
will lower the high energy cutoff scale. During inflation, the smallest strong coupling scale
actually is $\sqrt{2/3}\,M_{pl}$ \cite{Germani:2014hqa}, so there is no problem for the application
of the effective field theory. Naively, around a Minkowski background, there is an energy
scale $\Lambda_M=(M^2M_{pl})^{1/3}$ which is below the inflationary energy scale. However,
as pointed out in \cite{Germani:2011ua}, in a non-trivial background, the diagonalization of
the scalar-graviton system for the canonically normalized fields sets
the energy scale to be $\sqrt{2/3}\,M_{pl}$ \cite{Germani:2011ua,Germani:2014hqa}.

\section{Cosmological perturbations}

In this section, we derive the linear perturbation around the flat FRW background. For convenience, we choose
the uniform field gauge,
\begin{equation}
\label{eq13}
\delta \phi \left( {x,t} \right) = 0,\;\;{h_{ij}} = {a^2}\left( {(1 + 2\zeta  + 2{\zeta ^2}){\delta _{ij}} + {\gamma _{ij}} + \frac{1}{2}{\gamma _{il}}{\gamma _{lj}}} \right),
\end{equation}
where $\zeta$ and $\gamma_{ij}$ denote the scalar and tensor fluctuations respectively, the tensor perturbation satisfies
${\partial_i}{\gamma^{ij}} = 0$ and ${h^{ij}}{\gamma _{ij}} = 0$. Both $\zeta$ and $\gamma_{ij}$ are first order quantities
and we have expanded $\zeta$ and $h_{ij}$ to the second order. Since the scalar and tensor modes are decoupled, so we
consider the scalar perturbation first.

\subsection{Scalar perturbations}

For the scalar perturbations, we expand the lapse and shift functions to the first order as $N=1+N_1$ and $N_i=\partial_i\psi+ N_i^T$,
where ${\partial^i}N_i^T = 0$. Substituting the expansion for $N$ and $N_i$ into Eqs. \eqref{eq15} and \eqref{eq14}, we get the solutions \cite{Germani:2010ux,Germani:2011ua}
\begin{equation}
\label{eq16}
\begin{split}
N_1& = \frac{\dot \zeta }{\bar H},\quad \bar H = \frac{H(1 - 3 \Upsilon/2)}{1 - \Upsilon/2},\quad
\Upsilon=\frac{\dot\phi^2}{M^2 M_{pl}^2},\\
\psi & =  - \frac{\zeta}{\bar H} + \chi, \quad N^T_i=0,\\
\partial_i^2 \chi & = \frac{a^2\Sigma }{\bar H^2}\frac{\dot \zeta}{1 - \Upsilon/2},\quad
\Sigma  = \frac{\dot\phi^2}{2M_{pl}^2}\left[1  + \frac{3 H^2 (1 + 3\Upsilon/2)}{M^2 (1 - \Upsilon/2)} \right].
\end{split}
\end{equation}
By using the above solution (\ref{eq16}) and the background Eqs. (\ref{eq5})-(\ref{eq7}),
we expand the action (\ref{eq3}) to the second order of $\zeta$ and get \cite{Germani:2011ua}
\begin{equation}
\label{eq17}
S_{\zeta^2} = \int dt d^3x\, M_{pl}^2 a^3 \left\{\frac{\Sigma}{\bar H^2}\dot\zeta^2 -\frac{\theta_s}{a^2}(\partial_i \zeta)^2 \right\},
\end{equation}
where
\begin{equation}
\label{eq18}
\theta_s=\frac{1}{a} \frac{d}{dt}\left[\frac{a}{\bar H}\left(1 - \frac{\Upsilon}{2} \right) \right] - 1 - \frac{\Upsilon}{2}.
\end{equation}
By using the canonically normalized field $v = z \zeta$, where
\begin{equation}
\label{normvars}
z = a M_{pl} \frac{\sqrt{2\Sigma}}{\bar H},
\end{equation}
the action (\ref{eq17}) becomes
\begin{equation}
\label{eq20}
S_{\zeta ^2} = \int d^3x d\tau \frac{1}{2} \left[ v{'^2} - c_s^2 (\partial_i v)^2 + \frac{z''}{z} v^2 \right],
\end{equation}
where the conformal time $\tau$ is related to the coordinate time by $dt=a d\tau$, the prime denotes the derivative with respect to $\tau$,
and the effective sound speed is $c_s^2 = \bar H^2 \theta_s/\Sigma$.

In terms of the slow-roll parameters, we find
\begin{gather}
\label{slrparameq1}
\Upsilon\approx \frac{2F}{1+9F}\epsilon_v\left[1-\frac{4(1+6F)}{3(1+9F)}\epsilon_v+\frac{2}{3}\eta_v\right],\\
\label{slrparameq2}
\bar H \approx H\left(1-\frac{2F}{1+9F}\epsilon_v\right),\\
\label{slrparameq3}
\Sigma\approx \frac{1+3F}{1+9F}H^2\epsilon_v\left[1-\frac{4(1+9F+9F^2)}{3(1+3F)(1+9F)}\epsilon_v+\frac{2}{3}\eta_v\right],\\
\label{slrparameq4}
\theta_s=\frac{1+3F}{1+9F}\epsilon_v\left[1-\frac{4+30F+42F^2}{3(1+9F)(1+3F)}\epsilon_v+\frac{2}{3}\eta_v\right],\\
c_s^2=1-\frac{2F(1+7F)}{(1+3F)(1+9F)}\epsilon_v.
\end{gather}
In the GR limit, we recover the result $c_s^2=1$. In the high friction limit, we get
\begin{gather}
\label{sleq2}
\Upsilon\approx \frac{2}{3}\epsilon\left(1-\frac{8}{3}\epsilon+\frac{2}{3}\eta\right),\\
\label{sleq2a}
\bar H \approx  H\frac{1 - \epsilon}{1 - \epsilon/3}\approx H\left(1-\frac{2}{3}\epsilon \right),\\
\label{sleq3}
\Sigma \approx \frac{H^2\epsilon(1-8\epsilon/3+2\eta/3) (1 + \epsilon)}{1 - \epsilon/3}\approx H^2\epsilon\left(1-\frac{4}{3}\epsilon+\frac{2}{3}\eta \right),\\
\label{eq19}
\theta_s \approx \epsilon\left(1  - \frac{14}{9} \epsilon + \frac{2}{3}\eta\right),\\
\label{cseq1}
c_s^2 = \frac{\bar H^2 \theta_s}{\Sigma} \approx 1-\frac{14}{9}\epsilon.
\end{gather}
Note that this result about $c_s^2$ is different from that in \cite{Germani:2011ua} because they missed the second order corrections
due to $\theta_s$ in Eq. (\ref{eq19}) and $\Upsilon$ in (\ref{sleq2}).
Following the standard canonical quantization procedure,
we define the operator
\begin{equation}
\label{veq21}
\hat v(\tau ,\vec{x}) = \int \frac{d^3k}{( 2\pi)^3}\left[v_k(\tau)\hat a_k e^{i\vec{k} \cdot \vec{x}}+v_k^*(\tau)\hat a_k^\dag e^{-i\vec{k} \cdot \vec{x}}\right],
\end{equation}
where the operators satisfy the standard commutation relations
\begin{equation}
\label{veq22}
\begin{split}
\left[\hat a_k,\ \hat a_{k'}^\dag\right]=(2\pi)^3\delta^3(\vec{k}-\vec{k'}),\\
\left[\hat a_k,\ \hat a_{k'}\right]=\left[\hat a_k^\dag,\ \hat a_{k'}^\dag\right]=0,
\end{split}
\end{equation}
and the mode functions obey the normalization condition
\begin{equation}
\label{eq23}
v_k' v_k^* - v_k {v_k^*}' =  - i.
\end{equation}
The Bunch-Davis vacuum is defined by $\hat a_k|0\rangle=0$.
Varying the action (\ref{eq20}), we obtain the Mukhanov-Sasaki equation for the mode function $v_k(\tau)$,
\begin{equation}
\label{eq21}
{v_k}'' + \left( {c_s^2{k^2} - \frac{{z''}}{z}} \right){v_k} = 0.
\end{equation}
Inside the horizon in the past, as $aH/k\rightarrow 0$, the asymptotic solution
that satisfies the normalization condition (\ref{eq23}) is
\begin{equation}
\label{eq24}
v_k \to \frac{1}{\sqrt{2c_s k}}e^{- ic_s k \tau}.
\end{equation}
In order to solve the Mukhanov-Sasaki equation, we need to find the time derivative $z''/z$.
From the definition (\ref{normvars}), we find
\begin{equation}
\label{zvareq1}
\frac{z''}{z}\approx a^2 H^2 \left(2+\frac{2-3F}{1+3F}\epsilon_H+3\eta_H\right).
\end{equation}
Since
\begin{equation}
\label{eq25}
\frac{d}{d\tau}\left( \frac{1}{aH} \right) =  - 1 + \epsilon_H,
\end{equation}
and $H$ and $\epsilon_H$ change very slowly during inflation, so we obtain
\begin{equation}
\label{eq26}
aH \approx  - \frac{1}{(1 -\epsilon_H)\tau}.
\end{equation}
Substituting this result into Eq. (\ref{zvareq1}), we get
\begin{equation}
\label{zvareq2}
\frac{z''}{z}\approx \frac {1}{\tau^2}\left(2+\frac{9(1+4F)}{1+9F}\epsilon_v-3\eta_v\right)=\frac{\nu^2-1/4}{\tau^2},
\end{equation}
where
\begin{equation}
\label{zvareq3}
\nu\approx \frac{3}{2}+\frac{3(1+4F)}{1+9F}\epsilon_v-\eta_v.
\end{equation}
Combining Eqs. (\ref{zvareq2}) and
(\ref{eq21}), finally we obtain the equation,
\begin{equation}
\label{eq27}
{v_k}'' + \left( c_s^2k^2 - \frac{\nu ^2 - 1/4}{\tau ^2} \right)v_k = 0,
\end{equation}
Treating $\nu$ as a constant, the solution is
\begin{equation}
\label{eq28}
v_k\left( \tau  \right) = \sqrt \tau  \left[ c_1H_\nu ^{(1)}\left(  - c_sk\tau  \right) + c_2H_\nu ^{(2)}\left(  - c_sk\tau  \right) \right],
\end{equation}
where $H_\nu^{(1)}(x)$ and $H_\nu ^{(2)}(x)$ are the first and second Hankel function, respectively.
From the asymptotic condition (\ref{eq24}), we obtain $c_2=0$.
Outside the horizon, the Hankel function has the asymptotic form,
\begin{equation}
\label{eq29}
H_\nu ^{(1)}\left( x \ll 1 \right)\sim\sqrt {\frac{2}{\pi }} e^{ - i\frac{\pi }{2}} 2^{\nu  - \frac{3}{2}} \frac{\Gamma \left( \nu  \right)}{\Gamma \left( 3/2 \right)}x^{ - \nu },
\end{equation}
and the mode function
\begin{equation}
\label{eq30}
v_k = e^{i(\nu  - 1/2)\frac{\pi }{2}} 2^{\nu  - \frac{3}{2}} \frac{\Gamma \left( \nu  \right)}{\Gamma \left( 3/2 \right)}\frac{1}{\sqrt {2c_sk} }\left(  - c_sk\tau  \right)^{1/2 - \nu }\propto z.
\end{equation}
Thus the scalar perturbation outside the horizon is almost a constant,
\begin{equation}
\label{eq31}
|\zeta _k| = 2^{\nu  - \frac{5}{2}} \frac{\Gamma \left( \nu  \right)}{\Gamma \left( 3/2 \right)}
\frac{H(1-\epsilon_H)^{\nu-1/2}}{k^{3/2}M_{pl}\sqrt {c_s\theta_s} }\left( \frac{c_sk}{aH} \right)^{3/2 - \nu }.
\end{equation}
The power spectrum of $\zeta$ is defined by the two-point correlation function
\begin{equation}
\label{eq32}
\left\langle \hat \zeta( \tau ,\vec{k})\hat \zeta ( \tau ,\vec{k}') \right\rangle  = \frac{2\pi^2}{k^3}\delta ^3\left( \vec{k} - \vec{k}' \right)P_\zeta \left( k \right),
\end{equation}
so we get the power spectrum
\begin{equation}
\label{eq33}
\begin{split}
P_\zeta& = \frac{k^3}{2\pi ^2}\left| \zeta _k \right|^2 \approx 2^{2\nu  - 3}\left( \frac{\Gamma \left( \nu  \right)}{\Gamma \left( {3/2} \right)} \right)^2\frac{H^2(1-\epsilon_H)^{2\nu-1}}{8\pi ^2c_s\theta_s M_{pl}^2}\left( \frac{c_sk}{aH} \right)^{3 -2 \nu }\\
&\left.\approx \left[\frac{1+9F}{1+3F}+\frac{-2(1+9C)+3(1-24C)F}{3(1+3F)}\epsilon_v-\frac{2(1+9F)(1-3C)}{3(1+3F)}\eta_v\right]\frac{H^2}{8\pi^2 M_{pl}^2\epsilon_v}\right|_{c_sk=aH},
\end{split}
\end{equation}
where the constant $C=-2+\gamma+\ln2\approx -0.73$. In the limit $F=0$, we recover the standard GR result.
In the high friction limit $F\gg 1$, the scalar power spectrum becomes
\begin{equation}
\label{hfpseq1}
P_\zeta \approx \left[1+\left(\frac{1}{3}-8C\right)\epsilon-\left(\frac{2}{3}-2C\right)\eta\right] \frac{1}{2M_{pl}^2\, \epsilon}\left(\frac{H}{2\pi}\right)^2
\left( \frac{c_sk}{aH} \right)^{3 - 2\nu },
\end{equation}
Note that in order to derive the second order correction to the scalar spectral tilt,
we need to provide the first order correction to the amplitude of the power spectrum \cite{Stewart:1993bc,Gong:2001he}.

To the first order of approximation, using the relation $d\ln k = (1- \epsilon_H)Hdt$, the scalar spectral tilt is \cite{Tsujikawa:2012mk}
\begin{equation}
\label{snseq1}
n_s-1=3-2\nu=2\eta_v-\frac{6(1+4F)}{1+9F}\epsilon_v.
\end{equation}
In the GR limit, the standard result $n_s-1=2\eta_v-6\epsilon_v$ is recovered. In the high friction limit, we get $n_s-1=2\eta-8\epsilon$.
Therefore, for the same $n_s$, the slow-roll parameter $\epsilon$ can be smaller and the tensor-to-scalar
ratio can be smaller in the high friction limit. To the second order, we get
\begin{equation}\label{slrnseq2}
\begin{split}
n_s-1&= \left. \frac{d\ln P_\zeta }{d\ln k} \right|_{c_sk = aH}=-\frac{6(1+4F)}{1+9F}\epsilon_v+2\eta_v+\frac{2}{3}\eta_v^2+\left(\frac{2}{3}-2C\right)\xi^2\\
&\qquad +\frac{2[-1+8C+(-17+60C)F+12(-4+9C)F^2]}{(1+3F)(1+9F)}\epsilon_v\eta_v\\
&\qquad -\frac{2[5+36C+12(1+33C)F+18(-11+84C)F^2+27(-25+72C)F^3]}{3(1+3F)(1+9F)^2}\epsilon_v^2.
\end{split}
\end{equation}

In the high friction limit, the scalar spectral tilt is
\begin{equation}
\label{eq34}
n_s - 1 \approx  - 8\epsilon  + 2\eta  + \left( \frac{50}{3} - 48C \right)\epsilon ^2 + \left( 24C-\frac{32}{3}\right)\epsilon \eta +\frac{2}{3}\eta^2 + \left( \frac{2}{3} -2C\right)\xi^2,
\end{equation}
and the running of the scalar spectral index is
\begin{equation}
\label{eq42}
n_s'=\left. \frac{dn_s}{d\ln k} \right|_{c_sk = aH} =  - 48\epsilon^2 + 24\epsilon \eta  - 2\xi^2.
\end{equation}

\subsection{Gravitational wave}

Now we consider the tensor perturbation. Expanding the action to the second order of the tensor
perturbation $\gamma_{ij}$, we obtain the quadratic action \cite{Germani:2011ua}
\begin{equation}
\label{eq4-3}
S=\int{d^3xdt\frac{M_{pl}^2}{8}a^3\left[\left(1-\frac{\Upsilon}{2}\right)\dot\gamma_{ij}^2
-\frac{1}{a^2}\left(1+\frac{\Upsilon}{2}\right)(\partial_l\gamma_{ij})^2\right]}.
\end{equation}
With the symmetric traceless tensor $e_{ij}^s$ which satisfies the following relation
\begin{equation}
\label{gwpoleq1}
\sum_i e^s_{ii}=0,\quad \sum_{i,j} e^s_{ij}e^{s'}_{ij}=2\delta_{ss'},
\end{equation}
the tensor perturbation can be written as
\begin{equation}
\label{gwpoleq2}
\gamma_{ij}=\sum_{s=+,\times}e^s_{ij}\gamma^s.
\end{equation}
By using the canonical variable $u^s=z_t\gamma^s$, where
\begin{equation}
\label{gwpoleq3}
z_t=\frac{\sqrt{2}}{2}a M_{pl}\sqrt{1-\frac{\Upsilon}{2}},
\end{equation}
the quadratic action (\ref{eq4-3}) becomes
\begin{equation}
\label{eq4-4}
S = \sum_{s  =  + , \times }\int {d^3xd\tau \frac{1}{2}\left[ ({u^s}')^2 - c_t^2( \partial _i u^s )^2 +\frac{z''_t}{z_t}(u^s)^2 \right]},
\end{equation}
where
\begin{gather}
\label{cteq1}
c_t^2=\frac{1+\Upsilon/2}{1-\Upsilon/2}\approx 1+\frac{2F}{1+9F}\epsilon_v,\\
\label{zteq1}
\frac{z^{''}_t}{z_t}\approx a^2 H^2(2-\epsilon_H)\approx \frac{1}{\tau^2}\left[2+\frac{3(1+3F)}{1+9F}\epsilon_v\right].
\end{gather}
Due to the non-minimally derivative coupling, the speed of gravitational wave $c_t$ is a little larger than the speed of light.
In the GR limit, $c_t=1$. Using the canonical quantization,
\begin{equation}
\label{usquant1}
\hat u^s = \int \frac{d^3k}{( 2\pi)^3}\left[u^s_k(\tau)\hat a_k e^{i\vec{k} \cdot \vec{x}}+u^{s*}_k(\tau)\hat a_k^\dag e^{-i\vec{k} \cdot \vec{x}}\right],
\end{equation}
we obtain the equation for the mode function $u^s_k$,
\begin{equation}
\label{eq4-5}
{u_k^s}'' + \left( c_t^2k^2 - \frac{\mu ^2 - 1/4}{\tau ^2} \right)u^s_k = 0,
\end{equation}
where
\begin{equation}
\label{4-6}
\mu\approx \frac{3}{2}+\frac{1+3F}{1+9F}\epsilon_v.
\end{equation}
The solution is
\begin{equation}
\label{eq30g}
u^s_k = e^{i(\mu  - 1/2)\frac{\pi }{2}} 2^{\mu  - \frac{3}{2}} \frac{\Gamma \left( \mu  \right)}{\Gamma \left( 3/2 \right)}\frac{1}{\sqrt {2c_tk} }\left(  - c_tk\tau  \right)^{1/2 - \mu }.
\end{equation}
The power spectrum of gravitational wave is
\begin{equation}
\label{eq4-7}
\begin{split}
P_T&=\frac{k^3}{2\pi^2} |\gamma_{ij}|^2=\frac{k^3}{\pi^2} \sum_{s  =  + , \times } \left|\frac{u^s_k}{z_t}\right|^2\\
& \approx  2^{2\mu-3}\left(\frac{\Gamma(\mu)}{\Gamma(3/2)}\right)^2 \frac{8(1-\epsilon_H)^{2\mu-1}}{c_t M_{pl}^2 (1+\Upsilon/2)}
\left(\frac{H}{2\pi}\right)^2\left( \frac{c_t k}{aH}  \right)^{3 - 2\mu }\\
&\approx\left. \left[1-\frac{2(1+C)+2(4+3C)F}{1+9F}\epsilon_v\right]\frac{8}{M_{pl}^2}\left(\frac{H}{2\pi}\right)^2\right|_{c_t k=aH}.
\end{split}
\end{equation}
In the limit $F=0$, we recover the standard GR result.
In the high friction limit, the power spectrum of gravitational wave is
\begin{equation}
\label{hfpteq1}
P_T\approx \frac{8}{M_{pl}^2}\left(1-\frac{8}{3}\epsilon-2C\epsilon\right)\left(\frac{H}{2\pi}\right)^2\left( \frac{c_t k}{aH}  \right)^{3 - 2\mu }.
\end{equation}
Combining Eqs. \eqref{eq33} and \eqref{eq4-7}, we get the tensor to scalar ratio
\begin{equation}\label{t2sr}
r\approx 16\epsilon_v\left[\frac{1+3F}{1+9F}-\frac{(1+3F)[4(1-3C)+27(1-2C)F]}{3(1+9F)^2}\epsilon_v+\frac{2(1+3F)(1-3C)}{3(1+9F)}\eta_v\right].
\end{equation}
Note that there is an ambiguity in the above definition due to the difference between the effective sound speeds and the horizon exits for the tensor and scalar modes.
Because $c_t\ge c_s$, for the same mode $k$, the tensor mode exits the horizon later, so we should take $c_t k=aH$.
In the limit $F=0$, we recover the standard GR result.
In the high friction limit $F\gg 1$, we get
\begin{equation}
\label{eq4-2}
r = \frac{P_T}{P_\zeta} \approx 16\epsilon\left[1-(3-6C)\epsilon+\left(\frac{2}{3}-2C\right)\eta\right]\approx -8n_T.
\end{equation}

To the first order of approximation, the tensor spectral tilt is
\begin{equation}\label{tnteq1}
n_T=-\frac{2(1+3F)}{1+9F}\epsilon_v.
\end{equation}
In both the GR and the high friction limits, we get $n_T\approx -2\epsilon$ to the first order of approximation.
To the second order of approximation, the tensor spectral index is
\begin{equation}\label{slrnteq2}
\begin{split}
n_T &= \left. \frac{d\ln P_T}{d\ln k} \right|_{c_t k = aH} \\
&\approx
-\frac{2(1+3F)}{1+9F}\epsilon_v\left[1+\frac{11+12C+6(7+9C)F}{3(1+9F)}\epsilon_v-\frac{2(2+3C)}{3}\eta_v\right].
\end{split}
\end{equation}
In the high friction limit, the tensor spectral tilt is
\begin{equation}
\label{eq4-1}
n_T \approx  - 2\epsilon  - \left( \frac{28}{3} + 12C \right)\epsilon^2 + \left( \frac{8}{3} + 4C \right)\epsilon \eta,
\end{equation}
and the running of the tensor spectral index is
\begin{equation}
\label{eq4-1g}
n_T'=\left.\frac{d n_T}{d\ln k}\right|_{c_t k=aH}\approx -12\epsilon^2+4\epsilon\eta.
\end{equation}

\section{Inflationary Models}
Now let us apply the general results obtained in the previous section to the power law potential, the double well potential, the hilltop
inflation and the natural inflation.

\begin{figure}
\centerline{\includegraphics[width=0.6\textwidth]{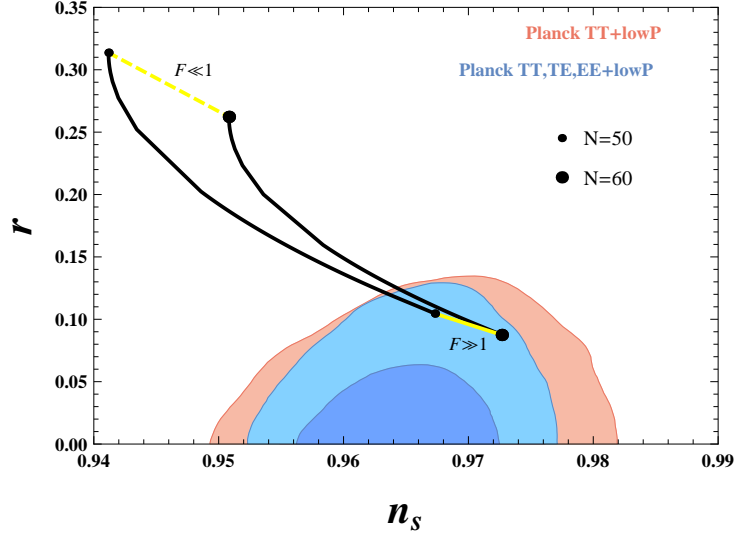}}
\caption{The theoretical results for the quartic potential $V(\phi)=\lambda\phi^4$ and the marginalized joint 68\% and 95\% CL regions for $n_s$ and $r_{0.002}$ from
Planck 2015 \cite{Ade:2015lrj}. The solid black lines show the effect of $F$.}
\label{higgsnsr}
\end{figure}

We use the quartic potential $\lambda\phi^4$ as an example to consider the effect of $F$ first. The results are shown in Fig. \ref{higgsnsr}.
As $F$ increases, the tensor to scalar ratio $r$ becomes smaller.
From Fig. \ref{higgsnsr}, we see that inflation with the quartic potential in the GR limit $F\ll 1$ is inconsistent with the Planck 2015 results, while the model
is consistent with the observation at the 2$\sigma$ level in the high friction limit $F\gg 1$. In the following discussion,
we consider inflationary models in the high friction limit only.

\subsection{Power Law Potential}
For the power-law potential
\begin{equation}
\label{eq43}
V\left( \phi  \right) = \lambda M_{pl}^4\left( \frac{\phi }{M_{pl}} \right)^n,
\end{equation}
the slow-roll parameters are
\begin{equation}
\label{pwrslreq1}
\begin{split}
\epsilon(\phi)&=\frac{n^2}{2\lambda}\frac{M^2 M_{pl}^n}{\phi^{n+2}},\\
\eta(\phi)&=\frac{n(n-1)}{\lambda}\frac{M^2 M_{pl}^n}{\phi^{n+2}},\\
\xi^2(\phi)&=\frac{n^2(n-1)(n-2)}{\lambda^2}\frac{M^4 M_{pl}^{2n}}{\phi^{2n+4}}.
\end{split}
\end{equation}
These formulae are also valid for the inverse power law potential.
For the inverse power law case with $n=-2$, all the above slow-roll parameters are constants.
However, inflation does not end for the intermediate inflation with inverse power law potential \cite{Barrow:1990vx,Muslimov:1990be}.

For $0<n<2$, inflation ends when $\epsilon(\phi_e)=1$, so
\begin{equation}
\label{pwrslreq2}
\phi_e=\left(\frac{n^2}{2\lambda}\right)^{1/(n+2)}(M^2 M_{pl}^n)^{1/(n+2)}.
\end{equation}
The number of e-folds before the end of inflation is
\begin{equation}
\label{pwrslreq3a}
N_*=\frac{\lambda\phi_*^{n+2}}{n(n+2)M^2 M_{pl}^n}-\frac{n}{2(n+2)}.
\end{equation}
For $n\ge 2$, inflation ends when $\eta(\phi_e)=1$, so
\begin{equation}
\label{pwrslreq2a}
\phi_e=\left(\frac{n(n-1)}{\lambda}\right)^{1/(n+2)}(M^2 M_{pl}^n)^{1/(n+2)},
\end{equation}
and the number of e-folds before the end of inflation is
\begin{equation}
\label{pwrslreq3b}
N_*=\frac{\lambda\phi_*^{n+2}}{n(n+2)M^2 M_{pl}^n}-\frac{n-1}{n+2}.
\end{equation}
So the value of scalar field at the horizon exit is
\begin{equation}
\label{pwrslreq6}
\phi_*=\left(\frac{n(n+2)\tilde N}{\lambda}\right)^{1/(n+2)}(M^2 M_{pl}^n)^{1/(n+2)},
\end{equation}
where $\tilde N=N_*+n/2(n+2)$ for $0<n<2$ and $\tilde N=N_*+(n-1)/(n+2)$ for $n\ge 2$.
In order to avoid quantum gravity, we require that $H^2\ll M_{pl}^2$,
this can be guaranteed if $\phi\ll M_{pl}$ during inflation. The sub-Planckian field excursion is possible
for the slow-roll inflation with the power-law potential because of the high friction condition $H^2\gg M^2$,
and the high friction limit is satisfied if $\phi_e\ll M_{pl}$.
For the Higgs inflation with $n=4$, the coupling constant $\lambda\approx 0.13$ at
the energy scale around $100$ GeV \cite{Agashe:2014kda},
if we ignore the running of the coupling constant, and take $N_*=60$ and $M=1.27\times 10^{-7} M_{pl}$,
then we find that $\phi_e=0.01M_{pl}$, $H(\phi_e)=5.1\times 10^{-5}M_{pl}$, $\phi_*=0.024 M_{pl}$
and the field excursion $\Delta\phi=0.014 M_{pl}$ which is sub-Planckian. If $\lambda$ is larger,
then the field excursion will be even smaller.

In terms of $N_*$, we get the value of the slow-roll parameters at the horizon exit $\phi_*$,
\begin{equation}
\label{eq44}
\epsilon  = \frac{n}{n + 2} \frac{1}{2\tilde N},\quad
\eta  = \frac{n - 1}{n + 2}\frac{1}{\tilde N},\quad
\xi^2  = \frac{(n - 1)(n - 2)}{(n + 2)^2}\frac{1}{\tilde N^2}.
\end{equation}
The values of the slow-roll parameters at the horizon exit do not depend on the model parameters $\lambda$ and $M$ explicitly,
and the results are a factor of $(n+2)/2$ smaller than the standard results in GR.
Because the second order slow-roll parameter $\xi^2$ is much smaller than the first order parameters, we consider the first order correction
only. The scalar spectral tilt is \cite{Tsujikawa:2012mk}
\begin{equation}
\label{eq45}
n_s - 1 =  - \frac{2(n+1)}{(n+2)\tilde N}.
\end{equation}
The running of the scalar spectral index is
\begin{equation}
\label{eq46}
n_s' =  - \frac{2\left( n + 1 \right)}{\left( n + 2 \right)\tilde N^2}.
\end{equation}
The tensor spectral tilt is $n_T=-2\epsilon=-n/(n+2)\tilde N$. The running of the tensor spectral index is
\begin{equation}
\label{eq45-1}
n_T' =  - \frac{n}{(n+2)\tilde N^2}.
\end{equation}
The tensor to scalar ratio is
\begin{equation}
\label{eq46-1}
r = \frac{8n}{(n+2)\tilde N}.
\end{equation}

\begin{figure}
\centerline{\includegraphics[width=0.8\textwidth]{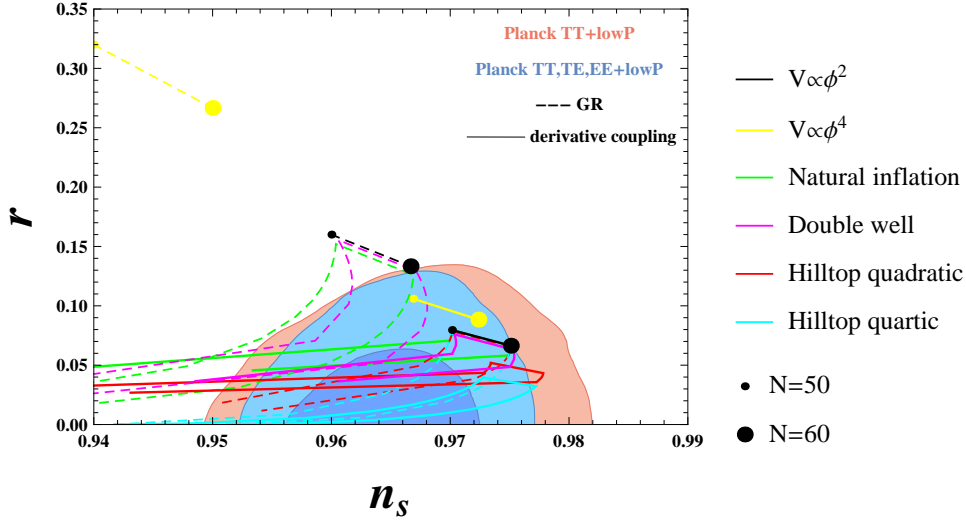}}
\caption{The theoretical results for some inflationary models and the marginalized joint 68\% and 95\% CL regions for $n_s$ and $r_{0.002}$ from
Planck 2015 \cite{Ade:2015lrj}. The solid lines are for the inflationary models with non-minimally derivative coupling and the
dashed lines are for the inflationary models in GR.}
\label{nsrpower}
\end{figure}

\begin{figure}
\centerline{\includegraphics[width=0.8\textwidth]{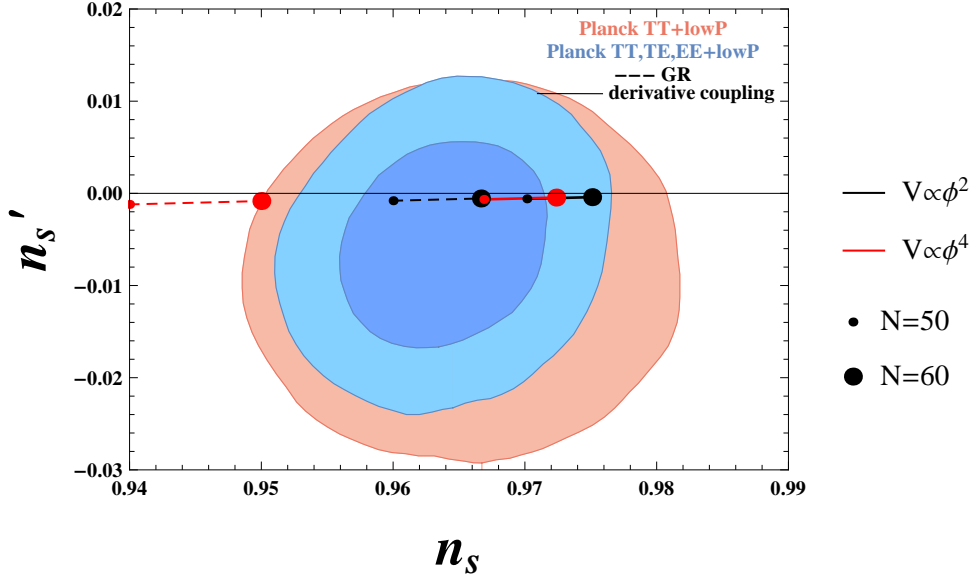}}
\caption{The theoretical results for the power law potentials with $n=2$ and $n=4$ and the marginalized joint 68\% and 95\% CL regions for $n_s$ and $n_s'$ from Planck 2015 \cite{Ade:2015lrj}. The solid lines are for the non-minimally derivative coupling and the
dashed lines are for GR.}
\label{nsdns2}
\end{figure}

The $n_s-r$ and $n_s-n_s'$ results for $n=2$ and $n=4$, along with the Planck 2015 constraints \cite{Ade:2015lrj}
are shown in Figs. \ref{nsrpower} and \ref{nsdns2}.
For comparison, we also plot the GR results \cite{Gong:2014cqa}.

\subsection{Hilltop models}

For the hilltop models with the potential \cite{Boubekeur:2005zm}
\begin{equation}
\label{hilltopeq1}
V(\phi)=\Lambda^4\left(1-\frac{\phi^p}{\mu^p}\right),
\end{equation}
the slow-roll parameters are
\begin{equation}
\label{slrhilltopeq1}
\begin{split}
\epsilon(\phi)&=\frac{ p^2 M^2 M_{pl}^4 (\phi/\mu)^{2 p-2}}{2 \Lambda ^4 \mu^2 \left[1-(\phi/\mu)^p\right]^3},\\
\eta(\phi)&=-\frac{ p (p-1) M^2 M_{pl}^4 (\phi/\mu)^{p-2}}{\Lambda ^4 \mu^2 \left[1-(\phi/\mu)^p\right]^2}
=-\frac{2(p-1)}{p}\frac{1-(\phi/\mu)^p}{(\phi/\mu)^p}\epsilon(\phi),\\
\xi^2(\phi)&=\frac{ p^2 (p-1) (p-2) M^4 M_{pl}^8(\phi/\mu)^{2 p-4}}{\Lambda ^8 \mu^4 \left[1-(\phi/\mu)^p\right]^4}.
\end{split}
\end{equation}
The high friction condition requires that
\begin{equation}
\label{hilltopfriceq1}
M^2 M_{pl}^2\ll \Lambda^4.
\end{equation}
The end of inflation is determined by
\begin{equation}
\label{phiehilltopeq1}
{\rm max}\{\epsilon(\phi_e),|\eta(\phi_e)|\}=1.
\end{equation}
For $p\ge 3$, $\epsilon(\phi)$ is smaller than $\eta(\phi)$ by a factor $(\phi/\mu)^p$. So the end of inflation is
determined by $|\eta(\phi_e)|=1$. If $\phi_e\ll \mu$ or $M^2 M_{pl}^4\gg \Lambda^4\mu^2$, we get
\begin{gather}
\label{phiehilltopeq2}
\left(\frac{\phi_e}{\mu}\right)^{p-2}\approx \frac{1}{p(p-1)}\frac{\Lambda^4}{M^2 M_{pl}^2}\left(\frac{\mu}{M_{pl}}\right)^2,\\
M^2M_{pl}^2\ll \Lambda^4\ll M^2M_{pl}^4/\mu^2.
\end{gather}
For the case $p=2$, $\eta(\phi)$ is almost a constant when $\phi\ll \mu$. Contrary to the cases with $p\ge 3$,
inflation happens when $M^2 M_{pl}^4\ll \Lambda^4\mu^2$
and inflation ends when $\phi_e\sim \mu$.

Because of the high friction condition,
the second order slow-roll parameter $\xi^2$ is smaller than the first order parameters $\epsilon$ and $\eta$
by a small factor $M^2 M_{pl}^2/\Lambda^4$, so we neglect the second order contribution.
The scalar spectral index is
\begin{equation}
\label{nshilltopeq1}
n_s-1=-\frac{2 p M^2 M_{pl}^4 x_*^{p-2} \left[(p-1)+(p+1) x_*^p\right]}{\Lambda ^4 \mu^2 \left[1-x_*^p\right]^3},
\end{equation}
where $x_*=\phi_*/\mu$.
The tensor to scalar ratio is
\begin{equation}
\label{rhilltopeq1}
r=\frac{8 p^2 M^2 M_{pl}^4 x_*^{2 p-2}}{ \Lambda ^4 \mu^2 \left(1-x_*^p\right)^3}.
\end{equation}
The number of e-folds before the end of inflation is $N_*=f(\phi_e/\mu)-f(\phi_*/\mu)$, where
\begin{equation}
\label{nhilltopeq1}
f(x)=\frac{\Lambda ^4 \mu ^2}{p M^2 M_{pl}^4}\left(\frac{x^{-p}}{2-p}+\frac{x^p}{p+2}-1\right)x^2,
\end{equation}
for $p\neq 2$. If $\phi\ll \mu$, we get
\begin{equation}
\label{nhilltopeq2}
N_*\approx \frac{\Lambda ^4 \mu ^2}{p(p-2) M^2 M_{pl}^4}\left(\frac{\mu}{\phi_*}\right)^{p-2}-\frac{p-1}{p-2},
\end{equation}
Therefore, the scalar spectral index can be written as
\begin{equation}
\label{nshilltopeq3}
n_s-1\approx -\frac{2(p-1)}{(p-2)[N_*+(p-1)/(p-2)]},
\end{equation}
and the scalar to tensor ratio is
\begin{equation}
\label{rhilltopeq3}
r\approx \frac{8px_*^p}{(p-2)[N_*+(p-1)/(p-2)]}\ll 1.
\end{equation}
If we take $p=4$, $\Lambda=\mu=0.01M_{pl}$ and $M=0.001\mu$, then we get $\phi_e=0.029\mu$,
$\phi_*=0.0045\mu$, $n_s=0.9512$, $n_s'=-0.0008$ and $r=1.07\times 10^{-10}$ for $N_*=60$.

For $p=2$, the function $f(x)$ is
\begin{equation}
\label{nhilltopeq3}
f(x)=\frac{\Lambda ^4 \mu ^2}{2 M^2 M_{pl}^4}\left(\ln x+\frac{x^4}{4}- x^2\right),
\end{equation}
and $\phi_*$ can be obtained from $N_*$ and $\phi_e$ by using the above function (\ref{nhilltopeq3}). For example,
if we take $\Lambda=10^{-2}M_{pl}$, $\mu=0.1M_{pl}$ and $M=10^{-6}M_{pl}$, we get $\phi_e=0.868\mu$ by setting
$\epsilon(\phi_e)=1$. For $N_*=60$, we find that $\phi_*=0.145\mu$ by using the function (\ref{nhilltopeq3}),
so $\Delta\phi=0.723\mu=0.0723M_{pl}$, $n_s=0.955$, $n_s'=-0.0002$ and $r=0.007$.

The $n_s-r$ results for $p=2$ and $p=4$ are shown in Fig. \ref{nsrpower}. In plotting the results for $p=4$,
we don't use the approximate relation (\ref{nshilltopeq3}) and (\ref{rhilltopeq3}), $\phi_e$ and $\phi_*$
are solved numerically instead.

\subsection{A simple symmetry breaking potential}

For the symmetry breaking potential \cite{Olive:1989nu}
\begin{equation}
\label{eq51}
V=\Lambda^4\left(1-\frac{\phi^2}{\mu^2}\right)^2,
\end{equation}
the slow-roll parameters are
\begin{equation}
\label{slrsymeq1}
\begin{split}
\epsilon(\phi)&=\frac{8M^2 M_{pl}^4(\phi/\mu)^2}{\Lambda^4\mu^2(1-\phi^2/\mu^2)^4},\\
\eta(\phi)&=-\frac{4M^2 M_{pl}^4(1-3\phi^2/\mu^2)}{\Lambda^4\mu^2(1-\phi^2/\mu^2)^4}
=-\frac{1}{2}\left(\frac{\mu^2}{\phi^2}-3\right)\epsilon(\phi),\\
\xi^2(\phi)&=-\frac{96M^4M_{pl}^8(\phi/\mu)^2}{\Lambda^8\mu^4(1-\phi^2/\mu^2)^7}.
\end{split}
\end{equation}
The potential is also called the double well potential.
In the region $\phi\gg \mu$, the above double well potential becomes the power
law potential with $n=4$. In the region $\phi\ll \mu$, the
above potential becomes the hilltop potential with $p=2$. Here we consider the intermediate region $\phi\sim \mu$,
this requires
\begin{equation}
\label{slrsymeq5}
M^2 M_{pl}^4\ll \Lambda^4\mu^2.
\end{equation}
In this region, $|\eta(\phi)|>\epsilon(\phi)$. Inflation ends when
\begin{equation}
\label{phiesymeq1}
\frac{(1-x_e^2)^4}{3x_e^2-1}=\frac{4M^2 M_{pl}^4}{\Lambda^4\mu^2},
\end{equation}
where $x_e=\phi_e/\mu$.
Note that the high friction condition requires that
\begin{equation}
\label{slrsymeq2}
\Lambda^4\left(1-\frac{\phi^2}{\mu^2}\right)^2\gg M^2 M_{pl}^2.
\end{equation}
The number of e-folds before the end of inflation is
\begin{equation}
\label{nsymeq1}
\begin{split}
N_*&=f(x_e)-f(x_*),\\
f(x)&=-\frac{\Lambda ^4 \mu^2}{4 M^2 M_{pl}^4}\left(\frac{x^6}{6}-\frac{3 x^4}{4}+\frac{3 x^2}{2}-\log (x)\right),
\end{split}
\end{equation}
where $x_*=\phi_*/\mu$.
The scalar spectral tilt is
\begin{equation}
\label{nssymeq1}
n_s-1=-\frac{8 M^2 M_{pl}^4 \left(1+x_*^2\right)}{\Lambda ^4 \mu^2 \left(1-x_*^2\right)^4}.
\end{equation}
The tensor to scalar ratio is
\begin{equation}
\label{rsymeq1}
r=\frac{128 M^2 M_{pl}^4x_*^2}{\Lambda^4\mu^2(1-x_*^2)^4}.
\end{equation}
For a given $N_*$, we can determine $\phi_*$ from Eqs. (\ref{phiesymeq1}) and (\ref{nsymeq1}).
If we take $\Lambda=10^{-2}M_{pl}$, $\mu=0.5M_{pl}$ and $M=10^{-6}M_{pl}$, we get $x_e=0.886$ from Eq. (\ref{phiesymeq1}).
For $N_*=60$, Eq. (\ref{nsymeq1}) gives $x_*=0.573$, so $\Delta\phi=(x_e-x_*)\mu=0.186 M_{pl}$, $n_s=0.975$, $n_s'=-0.0004$ and $r=0.046$.

The $n_s-r$ results in the intermediate region $\phi\sim \mu$ are shown in Fig. \ref{nsrpower}.

\subsection{Natural inflation}

For the natural inflation with the potential \cite{Freese:1990rb}
\begin{equation}
\label{nateq1}
V(\phi)=\Lambda^4\left[1+\cos\left(\frac{\phi}{f}\right)\right],
\end{equation}
the slow-roll parameters are
\begin{equation}
\label{eq60}
\begin{split}
\epsilon  = \frac{ M^2 M_{pl}^4\sin^2(\phi/f) }{2 f^2 \Lambda ^4 \left[1 + \cos(\phi/f) \right]^3},\\
\eta  =  - \frac{M^2 M_{pl}^4\cos(\phi/f)}{f^2 \Lambda ^4 \left[1 + \cos(\phi/f) \right]^2},\\
\xi^2 = -\frac{M^4 M_{pl}^8 \sin^2(\phi/f)}{f^4 \Lambda ^8 \left[\cos(\phi/f)+1\right]^4}.
\end{split}
\end{equation}
The high friction condition requires that
\begin{equation}
\label{highnatualeq1}
\Lambda^4\left[1+\cos(\phi/f)\right]\gg M^2 M_{pl}^2.
\end{equation}
So the second order slow-roll correction can be neglected. The scalar spectral index is
\begin{equation}
\label{nsnatural}
n_s-1=\frac{M^2 M_{pl}^4 \left[\cos(\phi_*/f)-2\right] \sec^4(\phi_*/2f)}{2 f^2 \Lambda ^4}.
\end{equation}
the horizon exit is determined by the number of e-folds before the end of inflation,
\begin{equation}
\label{eq62}
N_* = f(x_e)-f(x_*),
\end{equation}
where
\begin{equation}
\label{nnaturaleq1}
f(x)=\frac{f^2 \Lambda ^4}{M^2 M_{pl}^4} \left(\cos x+4 \ln\left[\sin\left(\frac{x}{2}\right)\right]\right),
\end{equation}
$x=\phi/f$, and the end of inflation is determined by
\begin{equation}
\label{slrendeq1}
\frac{\sin^2 x_e }{(1 + \cos x_e )^3}=\frac{2 f^2 \Lambda ^4}{M^2 M_{pl}^4}.
\end{equation}
If $f^2\Lambda^4/(M^2 M_{pl}^4)\gg 1$, then $x_e\sim \pi$ and we get \cite{Tsujikawa:2012mk}
\begin{equation}
\label{naturaleq6}
(\pi-x_e)^4\approx \frac{4 M^2 M_{pl}^4}{f^2\Lambda^4}.
\end{equation}
Since inflation happens around $\phi/f\sim \pi$, the potential behaves like the quadratic potential.
By solving Eqs. (\ref{eq62}) and (\ref{nnaturaleq1}), we obtain
\begin{equation}
\label{naturaleq7}
(\pi-x_*)^4\approx \frac{16 M^2 M_{pl}^4}{f^2\Lambda^4} \left(N_*+\frac{1}{4}\right).
\end{equation}
As we discussed above, the result is the same as the power law potential with $n=2$.
Substituting the above result (\ref{naturaleq7}) into Eq. (\ref{nsnatural}), we get \cite{Tsujikawa:2012mk}
\begin{gather}
\label{naturaleq8}
n_s-1\approx -\frac{6}{4N_*+1},\\
\label{naturaleq9}
r\approx \frac{16}{4N_*+1}.
\end{gather}
Note that the high friction condition (\ref{highnatualeq1}) requires that $\Lambda^2\gg fM$. If we take $\Lambda=0.01M_{pl}$,
$f=0.1M_{pl}$ and $M=10^{-6}M_{pl}$, then we find that $\phi_e=0.27 M_{pl}$, $\phi_*=0.14M_{pl}$,
$n_s=0.974$, $n_s'=-0.0004$ and $r=0.05$
for $N_*=60$.

If $f^2\Lambda^4\ll M^2 M_{pl}^4$, then either $\epsilon(\phi)$ or $\eta(\phi)$ is big for $0<\phi/f<\pi$,
so slow-roll inflation can not happen for this choice of model parameters.
The $n_s-r$ results along with the GR results \cite{Gong:2014cqa,Gao:2014yra}
are shown in Fig. \ref{nsrpower}. In plotting the results,
we don't use the approximate relation (\ref{naturaleq8}) and (\ref{naturaleq9}), $\phi_e$ and $\phi_*$
are solved numerically.

\section{Conclusions}
By introducing the slow-roll parameters defined in Eq. \eqref{slrparm1}, we obtain
the general expressions for the scalar and tensor spectral tilts to the second order
for the inflationary models with non-minimally derivative coupling. The results
can recover the well known GR results in the limit $H^2\ll M^2$. Furthermore, we extend the results of
the scalar and tensor spectral tilts to the second order in slow-roll parameters in the high friction limit $H^2\gg M^2$.
The non-minimal coupling of the kinetic term to Einstein tensor leads to enhanced friction for the scalar
field so that inflation happens more easily. The Lyth bound is modified with an extra suppression factor
$M/H$ so that the field excursion of the inflaton is sub-Planckian.

For the power law potential $V(\phi)\sim \phi^n$, due to the non-minimally derivative coupling, the
field excursion of the inflaton is sub-Planckian. The tensor to scalar ratio $r$ is a factor of $(n+2)/2$ smaller
than the result in GR which brings the quartic and quadratic potentials to be consistent with the observation at the 95\% CL.

For the hilltop potential, the scale $\mu$ can be smaller than the Planck energy. For the case $p=2$, inflation ends
when $\phi\sim \mu$. For $p>2$, small field inflation is realized, the tensor to scalar ratio $r$ is negligibly small,
and an approximate relation between $n_s$ and $N_*$ is derived in Eq. (\ref{nshilltopeq3}).

For the double well potential, the potential behaves like power law potential with $n=4$ in the regime $\phi\gg \mu$
and the hilltop potential with $p=2$ in the regime $\phi\ll \mu$. In the intermediate regime $\phi\sim \mu$,
we obtain $\Delta\phi=0.186 M_{pl}$, $n_s=0.975$ and $r=0.046$.

For natural inflation, inflation happens around the potential minimum $\phi/f\sim \pi$, and the behaviour
is similar to the quadratic potential $V(\phi)\sim \phi^2$. Due to the suppression of the non-minimally kinetic coupling, the
symmetry breaking scale $f$ can be smaller than the Planck energy.

In conclusion, the non-minimally kinetic coupling to Einstein tensor brings
the energy scale appeared in the hilltop inflation, double well potential and natural inflation
down to sub-Planckian scale, and the field excursion of the inflaton becomes sub-Planckian. The model is more
consistent with the observations in the high friction limit.

\begin{acknowledgments}

This research was supported in part by the Natural Science
Foundation of China under Grants No. 11175270, and No. 11475065;
the Program for New Century Excellent Talents in University under Grant No. NCET-12-0205;
and the Fundamental Research Funds for the Central Universities under Grant No. 2013YQ055.

\end{acknowledgments}


\providecommand{\newblock}{}

\end{document}